\documentclass[manuscript]{acmart}

\usepackage{subcaption}
\usepackage{enumitem}
\usepackage{multirow}
\usepackage{makecell}
\usepackage{graphicx}
\usepackage{tcolorbox}
\tcbuselibrary{skins, listings, breakable}
\setlist{nosep}

\AtBeginDocument{%
  }
\setcopyright{none}
\renewcommand\footnotetextcopyrightpermission[1]{}
\copyrightyear{2027}
\acmYear{2027}
\acmConference[CHI '27]{CHI Conference on Human Factors in Computing Systems}{May 10--14, 2027}{Pittsburgh, PA, USA}
\acmSubmissionID{9228}

\begin{document}

% \title{OmniScribble: From Single Sketched Objects to Living Scenes via LLM-Driven Animation}

\title{OmniScribble: Mixed-Initiative Authoring of Animated Sketches through Inspectable and Repairable AI Decisions}


\author{Hien Trinh}
\authornote{Both authors contributed equally to this research.}
\orcid{0009-0006-0396-7219}
\affiliation{%
  \institution{University of Science}
  \city{Ho Chi Minh}
  \country{Vietnam}
}
\affiliation{%
  \institution{Vietnam National University}
  \city{Ho Chi Minh}
  \country{Vietnam}
}

\author{Tan-Hiep To}
\authornotemark[1]
\orcid{0009-0008-1264-1354}
\affiliation{%
  \institution{University of Science}
  \city{Ho Chi Minh}
  \country{Vietnam}
}
\affiliation{%
  \institution{Vietnam National University}
  \city{Ho Chi Minh}
  \country{Vietnam}
}

\author{Trung-Nghia Le}
\orcid{0000-0002-7363-2610}
\affiliation{%
  \institution{University of Science}
  \city{Ho Chi Minh}
  \country{Vietnam}
}
\affiliation{%
  \institution{Vietnam National University}
  \city{Ho Chi Minh}
  \country{Vietnam}
}
\authornote{Corresponding author. Email: ltnghia@fit.hcmus.edu.vn}
\renewcommand{\shortauthors}{Trovato et al.}

\begin{abstract}

 AI-assisted creative tools reduce the effort of authoring content, but delegating creative decisions to AI can make it difficult for authors to understand, correct, or redirect work. We present OmniScribble, a system for animating multi-object sketches that treats object decomposition and motion planning as editable artifacts. Rather than generating animation end-to-end, OmniScribble presents AI proposals at verification checkpoints, where authors can accept, repair, or regenerate them. Both manual and AI-assisted authoring operate on the intermediate representation and rendering pipeline, allowing us to isolate AI delegation. In a within-subject study with eight participants, AI-assisted authoring yielded a median within-participant reduction in completion time of 71.4\%, while perceived control and motion-intent alignment remained high. Interaction traces and interviews show that participants adapted to AI errors by repairing local issues, regenerating mismatches, and accepting proposals. These findings suggest that inspectable, repairable delegation enables generative automation while maintaining control over the outcome.

\end{abstract}



\begin{teaserfigure}
  \includegraphics[width=\textwidth]{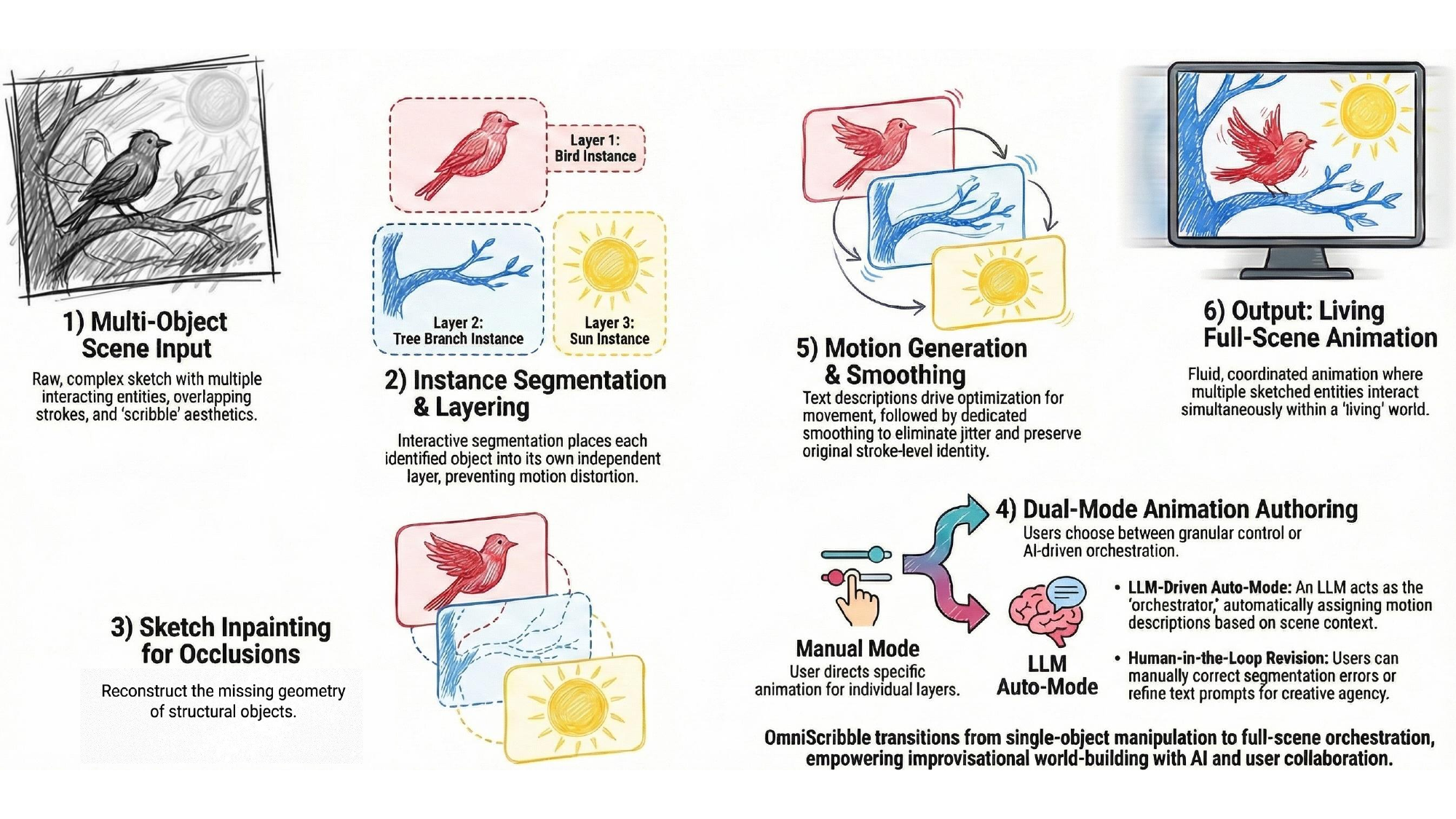}
  \caption{
  OmniScribble conceptual workflow. The system supports mixed-initiative sketch-animation authoring by exposing AI-generated intermediate decisions for user inspection and intervention. Users can construct or delegate object decomposition and motion planning, then accept, directly edit, or regenerate AI proposals at verification checkpoints. Both authoring modes share the same editable representations and rendering pipeline, enabling users to flexibly balance AI assistance with direct creative control.
  }
  \Description{Conceptual overview of OmniScribble. A multi-object sketch is first decomposed into separate object instances and layers, followed by sketch inpainting for occluded regions. Users can then author object-specific motion either manually or with AI assistance. In Auto mode, AI-generated object decomposition and motion proposals are presented at verification checkpoints, where users can accept, manually edit, or regenerate them. The resulting object layers and trajectories are passed to a shared animation-synthesis pipeline that produces a full-scene animation.}
  \label{fig:teaser}
\end{teaserfigure}


\maketitle

%=========================

\section{Introduction}
Imagine sketching a scene with a bird perched on a tree branch, the sun overhead, and wind moving through the scene, as shown in Fig.\ref{fig:teaser}. Turning this simple sketch into an animation requires more than generating plausible motion. The system must first determine which strokes form each semantic object (\textit{e.g.}, the bird, branch, sun) and then determine how each object should move. For example, the bird may flap its wings and take flight, creating a gust of wind that makes the branch sway; the sun may remain stationary. These motions are object-specific and must be coordinated over time. Errors in either step can cause the animation to diverge from the author's intent, such as assigning strokes to the wrong object or applying an inappropriate trajectory. This highlights a fundamental challenge in AI-assisted sketch animation: generating plausible motion is only part of the authoring process; authors must also be able to inspect and steer the object decomposition and object-specific trajectories that shape the result.

Sketches offer a lightweight and expressive way to convey objects, actions, and interactions without specifying every detail. Recent advances in generative video models~\cite{guo2023animatediff,lin2024animatediff} and vector-graphics animation~\cite{ma2022towards,jain2023vectorfusion,smith2023method} have made sketch-to-animation increasingly accessible. For multi-object scenes, however, animation requires more than generating motion: systems must identify object boundaries and stroke ownership, reason about interactions, and coordinate object-specific trajectories. Recent pipelines automate these steps, including object decomposition, occlusion reconstruction, trajectory planning, and motion synthesis~\cite{liu2025multi}. While this reduces authoring effort, these intermediate decisions are often hidden from authors, making it difficult to understand or correct mismatches with their intent.

Existing animation tools largely lie between direct manipulation and end-to-end generation. Direct manipulation provides precise control but makes multi-object authoring tedious, while end-to-end generation reduces effort but hides many intermediate decisions. A less explored alternative is to introduce AI assistance as a sequence of inspectable and correctable steps~\cite{chung2022talebrush,suh2023sensecape,yang2026storyspark}. This allows authors to delegate individual decisions to AI while retaining the ability to review and revise them when needed. We explore this design space by treating object decomposition and object-specific trajectories as shared, editable artifacts that can be created manually or delegated to AI.

We present \textbf{OmniScribble}, a mixed-initiative sketch-animation interface that supports both manual authoring and step-wise AI assistance. Rather than treating generation as a single operation, OmniScribble represents key intermediate decisions, object decomposition and trajectory specification, as explicit artifacts that can be inspected and edited. The workflow starts from a vector sketch and a motion caption. In Manual mode, users construct these artifacts directly. In Auto mode, the system proposes them sequentially and pauses at checkpoints for user intervention. At each checkpoint, users can accept the proposal, repair it directly, or regenerate it, optionally providing natural-language feedback. Once the specification is finalized, the system renders the animation using our diffusion-guided multi-object pipeline, which synthesizes coordinated rigid and non-rigid stroke motion while preserving each object's original vector identity.

This design lets us study AI assistance as delegation of individual authoring decisions, rather than as a binary choice between manual work and full automation. We conduct a within-subject study in which participants create sketch animations in both Manual and Auto modes across tasks of varying complexity. We examine how AI assistance affects authoring efficiency, workload, perceived control, and alignment with author intent. More importantly, we examine how authors respond when AI proposals are imperfect: when do they repair a problem themselves, when do they ask AI to regenerate, and when do they simply accept the proposal? These behaviors reveal how authors allocate correction work between themselves and AI during an iterative creative workflow.

Our contributions are as follows:
\begin{itemize}

    \item We introduce OmniScribble, an interactive sketch-animation system that represents object decomposition and trajectory specification as shared, editable intermediate artifacts. These artifacts can be created by users or proposed by AI, supporting different levels of automation while keeping intermediate decisions visible.
    
    \item We introduce a checkpoint-driven interaction model for AI-assisted sketch animation, in which users can accept, directly repair, or regenerate each AI proposal with optional natural-language guidance. This model moves beyond binary accept-or-reject automation by allowing authors to decide whether to repair a proposal themselves or delegate the correction back to AI.
    
    \item We characterize how authors respond to imperfect AI proposals, showing how they selectively allocate correction work between direct manipulation and AI regeneration, alongside effects on authoring efficiency, workload, perceived control, and outcome alignment.

\end{itemize}

\section{Related Work}
\subsection{Sketch Animation and Authoring Control}

Sketch animation has traditionally relied on direct manipulation, where authors manually group strokes, define movable layers, and specify motion through keyframes or trajectories. These approaches provide precise control but become increasingly laborious as the number of objects and interactions grows. Recent generative approaches reduce this burden by leveraging pretrained motion and video models. Live-Sketch~\cite{Gal_2024_CVPR} and FlipSketch~\cite{bandyopadhyay2025flipsketch} animate single-object sketches without manual keyframing, while multi-object systems introduce additional decisions about stroke ownership and object-level motion.

GroupSketch~\cite{liang2025multi} retains user involvement by asking authors to group strokes and initialize trajectories, whereas MoSketch~\cite{liu2025multi} automates object decomposition and motion planning to reduce manual effort. Building upon prior work, our system similarly automates multi-object animation, including occlusion-aware reconstruction and diffusion-guided motion optimization. These approaches demonstrate the benefit of automation, but differ in how much of the intermediate authoring process remains accessible to users. In particular, object decomposition and trajectory planning are typically either specified manually or inferred automatically, leaving limited opportunity to inspect and revise these decisions before they affect the final animation.

Vector-graphics systems further explore structured representations for editable and language-based animation. Keyframer~\cite{tseng2024keyframer} and Decomate~\cite{park2025decomate} use multimodal language models to interpret structured graphics and translate natural-language instructions into editable animation code, while InternSVG~\cite{wang2026internsvg} and VectorPrism~\cite{yun2026vector} support semantic grouping, editing, and animation. These systems highlight the value of structured representations for creative control, but primarily focus on manipulating the representation or final animation rather than exposing the intermediate decisions involved in decomposing and planning motion for a multi-object sketch.

\subsection{Structured and Mixed-Initiative Human--AI Co-Creation}

Recent HCI research frames generative AI as a collaborator embedded within iterative authoring workflows rather than as an isolated content generator. Mixed-initiative systems distribute contributions across
human and automated agents, enabling generated outputs to serve as working material for inspection, editing, and subsequent exploration across domains such as programming, visual design, and music composition~\cite{chiou2023designing,shin2024facilitating,gross2024towards}.
An increasing line of work demonstrates that effective collaboration requires interaction mechanisms beyond conversational prompting: IntentTagger~\cite{gmeiner2025intent} decomposes user goals into modular, independently adjustable units, PromptCharm~\cite{wang2024promptcharm} integrates localized visual refinements to align generated images with user expectations, and VideoDiff~\cite{huh2025videodiff} supports parallel generation and comparison during video editing.

To coordinate such collaboration, interactive systems increasingly depend on structured intermediate representations. Prior investigations highlight compositional substrates as essential frameworks for maintaining spatial and conceptual orientation during generation~\cite{cao2025compositional}. A similar pattern appears across recent systems: Luminate~\cite{suh2024luminate} organizes latent spaces into structured alternatives, while  Flowco~\cite{freund2025flowco}, ImaginationVellum~\cite{marquardt2025imaginationvellum}, and PaperBridge~\cite{zhang2025paperbridge} demonstrate that externalizing creative states through dataflow graphs, spatial canvases, and narrative scaffolds provides tangible artifacts for mutual intervention. These shared representations transition the interaction paradigm from pure text prompting toward direct manipulation of persistent structures accessible to both users and models.

The specific distribution of responsibility across these structures directly influences authorial agency and cognitive load. Empirical findings demonstrate that varying levels of automated assistance change mental allocation, confirming that co-led workflows with opportunities for manual revision foster stronger process ownership~\cite{xu2025productive}. OmniScribble builds upon these foundations by exposing two consequential stages of animation synthesis: object decomposition and trajectory choreography. By rendering both stages as inspectable, editable intermediate representations, the system establishes a clear operational surface where authors can seamlessly evaluate, repair, or delegate tasks between themselves and the model.

\subsection{Agency, Control, and Selective Delegation}

Determining appropriate system agency while preserving user influence represents a core tension in human--AI co-creation. In creative drawing, users perceive automated assistants as passive tools or active collaborators depending on how system initiative manifests across the interaction~\cite{lawton2023tool}. Consequently, agency is co-constructed through ongoing collaboration rather than assigned exclusively to either partner~\cite{luo2024co}.

Control is likewise not necessarily established once at the beginning of an authoring session. Recent work conceptualizes control as something that evolves over the course of human--AI co-creation~\cite{issak2026control}, while systems for regulating AI participation demonstrate the value of allowing users to determine when and how automated agents intervene
\cite{houde2025controlling}. Furthermore, uncritical cognitive offloading risks blinding creators to subtle system errors, underscoring the need for interfaces that stimulate reflection and rigorous evaluation before intermediate outputs propagate downstream~\cite{gmeiner2025exploring}.

Effective intervention requires functional mechanisms to evaluate, preserve, and revise machine proposals. Prior systems support this balance using direct manipulation, persistent histories, and structured variations~\cite{cao2025compositional,houde2025controlling,lawton2023tool,marquardt2025imaginationvellum}. OmniScribble instantiates this principle through selective delegation over inspectable intermediate representations. At each checkpoint, authors can accept a proposal, repair the shared artifact via direct manipulation, or delegate revisions back to the model through language-guided regeneration. This mechanism allows responsibility for the same artifact to shift fluidly between user and system, enabling our study to evaluate how creators distribute corrective effort when resolving machine errors.

\section{Design Rationale and Goals}
\label{sec:rationale}

OmniScribble is designed around a simple tension in AI-assisted creative authoring: automation can reduce effort, but opaque automation can make important decisions difficult to understand and change. Rather than treating AI assistance as an all-or-nothing choice, we explore a workflow in which authors and AI can share responsibility for individual authoring decisions. This requires making those decisions visible, keeping them editable, and allowing authors to choose how much assistance they want at each stage.

\paragraph{Expose consequential intermediate decisions.}
Multi-object sketch animation involves two closely related decisions: \emph{what} should move and \emph{how} it should move. Object decomposition determines which strokes belong together, while object-specific trajectories determine how those objects behave over time. In an end-to-end generation pipeline, both decisions may be hidden inside the generation process, making it difficult for authors to understand why an animation differs from their intent. We therefore expose object decomposition and trajectory specification as explicit intermediate artifacts that can be inspected before final rendering.

\paragraph{Use a shared representation for manual and AI-assisted authoring.}
To support both manual authoring and AI assistance within the same workflow, we use a shared representation consisting of an object decomposition and a trajectory for each object. Users and AI assistance can therefore contribute to the same underlying artifacts rather than working through separate workflows. Regardless of how the artifacts are created, they are passed to the same rendering pipeline. This keeps the distinction between Manual and Auto modes focused on who creates the intermediate decisions, rather than introducing differences in the final animation process.

\paragraph{Support granular rather than monolithic delegation.}
Authors may want to delegate some decisions while retaining direct control over others. Moreover, the usefulness of AI assistance may vary across authoring stages and individual proposals. We therefore treat automation as stage-wise delegation rather than a single decision to automate the entire process. At each stage, an AI-generated artifact is presented as a proposal that authors can inspect before deciding whether to continue, intervene directly, or request further assistance.

\paragraph{Make imperfect AI proposals repairable rather than disposable.}
AI proposals are rarely simply correct or unusable. A proposal may capture the intended motion while containing a small stroke-assignment error or an imprecise trajectory. Requiring authors to either accept or discard such a proposal wastes useful work. We therefore support three responses at each checkpoint: \emph{accept} the proposal, \emph{repair} it directly, or \emph{regenerate} it with AI assistance. This allows authors to choose between direct manipulation and further delegation based on the nature of the problem.

\section{OmniScribble: Mixed-Initiative Interface for Sketch-Animation Authoring}

We instantiate these design goals in OmniScribble, a mixed-initiative interface for multi-object sketch-animation authoring. Rather than treating animation as a single generation operation, OmniScribble organizes the authoring process into a sequence of explicit stages. Each stage produces an intermediate representation that users can create themselves or obtain through system assistance. AI-generated representations remain editable, allowing users to inspect the proposal, make corrections, and decide when to move forward with the authoring process. This structure divides work between the author and the system explicitly while preserving opportunities for correction throughout the workflow.

\subsection{Interface Design}

\begin{figure}[t!]
    \centering
    \includegraphics[width=\columnwidth]{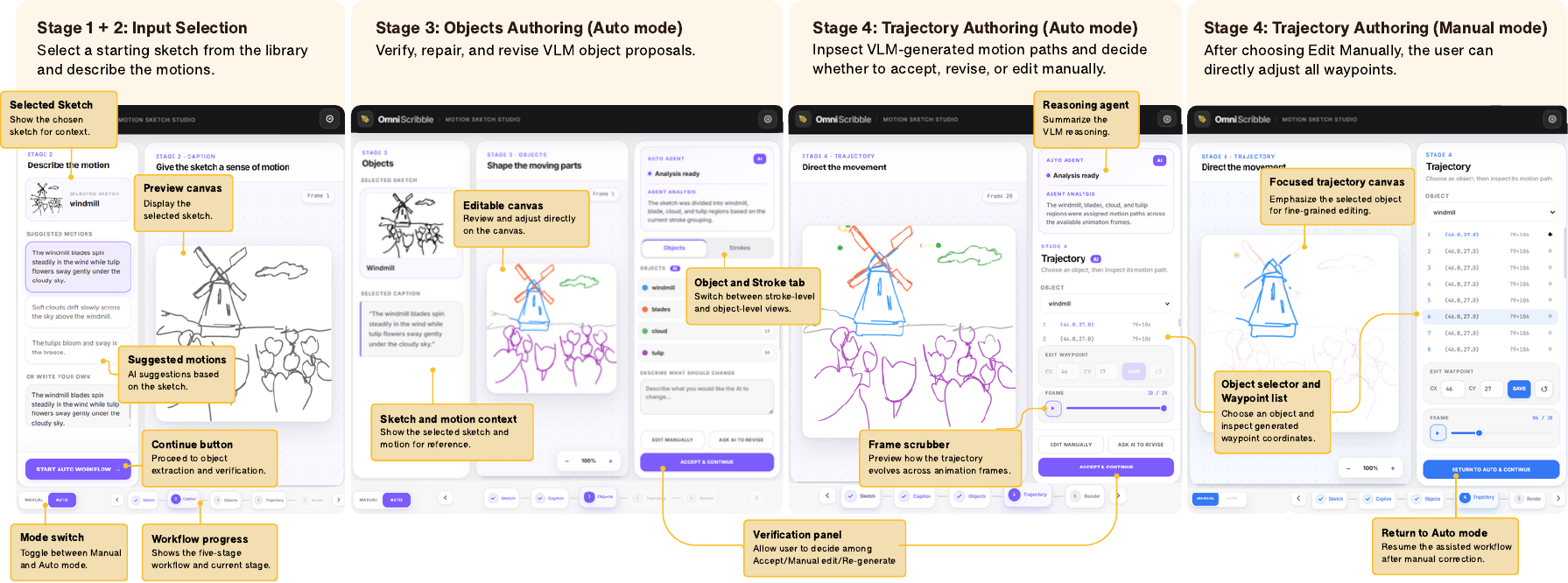}
    % \vspace{-5mm}
    \caption{Overview of the OmniScribble interface across representative authoring stages. From left to right: (a) sketch and motion selection, where users choose an input sketch and specify the intended motion; (b) object authoring in Auto mode, where VLM-proposed object groups are exposed for inspection and correction; (c) trajectory authoring in Auto mode, where users review generated motion paths and may accept, manually edit, or request revision; and (d) trajectory authoring after switching to Manual mode, where individual waypoints can be directly adjusted before returning to the assisted workflow.}
    \Description{Overview of the OmniScribble interface across four representative stages. From left to right, the panels show: selecting a sketch and describing its intended motion; reviewing AI-generated object groups in Auto mode; inspecting AI-generated object trajectories with a verification panel and frame scrubber; and manually editing trajectory waypoints after switching from Auto to Manual mode. The interface maintains the original sketch and motion description as persistent context and provides editable canvases for object and trajectory manipulation.}
    \label{fig:interface}
\end{figure}

OmniScribble structures authoring workflow into a five-stage progression: Sketch, Caption, Objects, Trajectory, and Render, maintaining a unified visual layout across Manual and Auto modes (Fig.~\ref{fig:interface}). A persistent stage navigator tracks overall progress, while a mode toggle clarifies whether the user or the system is responsible for constructing intermediate representations. 

\paragraph{Persistent authoring context.}
The interface preserves the selected sketch and motion prompt in a dedicated panel throughout authoring. Keeping these source materials visible grounds subsequent object grouping and trajectory authoring in the original visual context, simplifying comparison between author intent and intermediate proposals.

\paragraph{Editable canvas as a shared interaction surface.}
The central canvas functions as the primary workspace for both authoring modes. In the Objects stage, semantic groupings appear directly over the drawing using distinct colors, allowing instant inspection and direct stroke-level edits. In the Trajectory stage, motion paths overlay the sketch alongside interactive waypoints and a frame scrubber, connecting spatial path editing with temporal previews. AI proposals are rendered as editable elements rather than static previews.

\paragraph{Verification checkpoints and direct repair.}
In Auto mode, generation pauses at explicit verification states after the Objects and Trajectory stages, exposing the proposal alongside a summary of the model's interpretation. Users select from three actions: \emph{Accept} to continue delegation, \emph{Edit Manually} to reclaim direct control, or \emph{Ask AI to Revise} to request regeneration. Selecting \emph{Edit Manually} preserves the existing AI proposal, highlights the target entity, and reveals direct editing tools to adjust strokes or waypoints without restarting from scratch. Once repairs are complete, users resume the automated pipeline, ensuring fluid interoperability between human intervention and automated generation on a shared representation.

\subsection{Shared Authoring Workflow}

\begin{figure}[t!]
    \centering
    \includegraphics[width=\columnwidth]{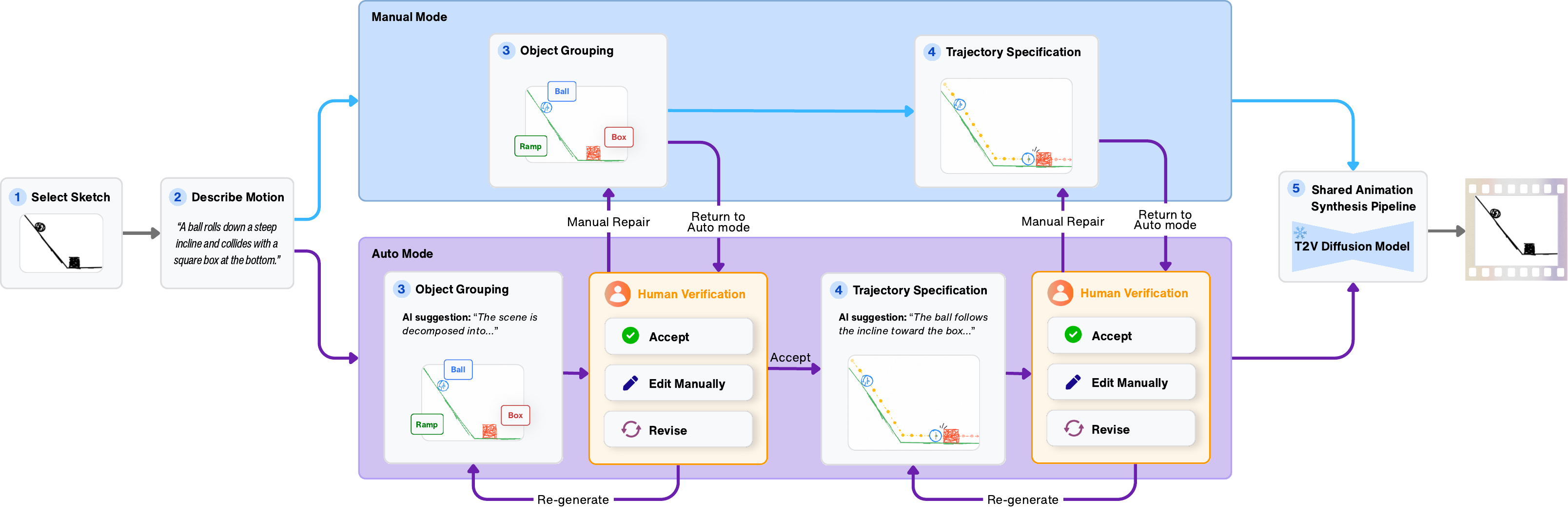}
    % \vspace{-5mm}
    \caption{OmniScribble's mixed-initiative authoring workflow. Following shared sketch selection and motion specification, users can directly construct object groups and trajectories in Manual mode or work from AI-generated proposals in Auto mode. Each AI proposal is exposed at a human verification checkpoint, where users may accept, manually repair, or request revision. Users can move between AI assistance and direct manipulation without discarding the current intermediate representation, and both paths converge on the same shared animation synthesis pipeline.}
    \Description{Workflow diagram comparing Manual and Auto authoring in OmniScribble. Both modes begin with sketch selection and motion description. In Manual mode, users construct object groups and trajectories directly. In Auto mode, the system generates these intermediate representations sequentially and pauses at human-verification checkpoints. At each checkpoint, users can accept the proposal, repair it manually, or regenerate it with AI assistance. Both authoring paths produce the same object and trajectory representations and converge on a shared animation-synthesis pipeline.}
    \label{fig:workflow}
\end{figure}

OmniScribble separates the authoring process into five stages: sketch selection, motion description, object construction, trajectory specification, and rendering (Tab.~\ref{fig:workflow}). The first two establish the author's initial intent, while the remaining steps construct the intermediate representations used to generate the final animation. A session begins with the user selecting a vector sketch from a set of candidate samples and specifying the intended motion through a system-suggested or custom caption. These steps are identical in Manual and Auto modes, ensuring that both conditions start from the same sketch and motion intent. Both modes begin with the same sketch and motion intent, but differ in how the subsequent authoring steps are performed.

In Manual mode, the user performs the remaining authoring steps directly, assigning sketch strokes to semantic objects, specifying a trajectory for each object, and initiating rendering without AI assistance.

In Auto mode, the user initiates automatic generation after selecting a caption. The system first proposes an object decomposition and pauses for verification. Once the object decomposition is confirmed, the system generates a trajectory for each object and presents the result for user review. The user can then accept, directly edit, or regenerate each proposal, as described in Section~\ref{sec:verify}. Thus, automation is introduced at individual authoring stages rather than as a single end-to-end generation operation.

Despite these different authoring paths, both modes produce the same intermediate representations before rendering: object layers with assigned strokes and per-object motion trajectories. The same rendering backend consumes these representations in both conditions. This shared representation isolates the primary difference between the modes to how the intermediate authoring decisions are produced: through direct manipulation in Manual mode or through AI-assisted proposals and user intervention in Auto mode.

\subsection{Manual Authoring}

Manual mode provides direct manipulation of the intermediate representations used by the animation pipeline. After selecting the sketch and motion description, users construct semantic objects by creating object groups and assigning the corresponding sketch strokes. Stroke assignments can be adjusted directly on the canvas, while unassigned strokes remain part of the static background.

Once the object structure is established, users specify an object-specific trajectory by drawing a path and positioning editable waypoints on the canvas. The trajectory can be inspected relative to the original sketch and adjusted before rendering.

Manual mode therefore makes the author responsible for both consequential intermediate decisions: determining which strokes constitute each movable object and specifying how each object should move. Importantly, these decisions are stored in exactly the same representation used in Auto mode. The difference between the two conditions is consequently \textit{who initially constructs the representation}, rather than how the resulting animation is synthesized.

\subsection{AI-Assisted Authoring}

Auto mode delegates the construction of intermediate representations to the AI while retaining the user as the decision maker at key points in the workflow. Rather than generating the complete animation in a single operation, OmniScribble introduces assistance sequentially.

After the user specifies the motion, the system first proposes an object decomposition. The proposal identifies semantic objects and associates the corresponding sketch strokes with each object. The result is displayed directly on the canvas using the same representation used in Manual mode, allowing users to inspect object boundaries and stroke ownership before any motion is generated.

Once the object representation has been verified, the system generates object-specific trajectories conditioned on the motion description and the verified objects. The resulting paths are displayed as editable overlays, allowing users to assess the proposed choreography before rendering.

This interaction changes the author’s role from constructing every intermediate representation to evaluating and selectively intervening in AI-generated proposals. Importantly, delegation does not remove authors from the workflow. Instead, it moves their effort toward deciding whether a proposal is sufficiently aligned with their intent and determining how any mismatch should be addressed. This distinction is central to OmniScribble’s mixed-initiative design: the AI performs construction, while the author retains responsibility for deciding whether and how that construction should proceed.

\subsection{Verification and Repair}
\label{sec:verify}

A central design goal of OmniScribble is to avoid treating automation as a single, all-or-nothing generation step. AI proposals can be neither fully correct nor entirely unusable: a result may capture the intended structure or motion while containing a localized error. Discarding such a proposal wastes useful work, while accepting it unchanged can carry the error into subsequent stages. OmniScribble therefore introduces a \emph{verification checkpoint} after each AI-assisted stage, where users can choose among three responses: \emph{accept} the proposal, \emph{repair} it directly, or \emph{regenerate} it with AI assistance. Together, these responses let users decide when to accept an AI proposal, when to intervene directly, and when to ask the AI to try again.

% \paragraph{Accept.}
% When the proposal matches the user's intended specification, the user approves it without modification, delegating the construction entirely to the AI.

% \paragraph{Direct repair.}
% When the proposal is largely usable but contains a localized mismatch, the user refines it directly using manual tools: adjusting stroke groupings for objects or moving waypoints for trajectories.

% \paragraph{AI regeneration.}
% When users judge a proposal to be unsuitable for local repair, the user requests a new proposal, optionally attaching concise natural-language feedback to guide the subsequent generation attempt.

These checkpoints turn verification into a flexible decision point across stages, enabling independent allocation of manual and automated effort for object grouping and motion choreography. OmniScribble records acceptance, manual repair, and regeneration actions together with their associated stage and timestamp. These interaction traces allow us to examine not only whether AI assistance reduces authoring effort, but also \emph{how users choose to allocate correction work between direct manipulation and AI assistance when generated intermediate results do not fully match their intent}.

\subsection{Animation Synthesis}
\label{sec:backend}

After the intermediate specification has been finalized, OmniScribble passes the object decomposition and object-specific trajectories to a shared animation synthesis pipeline. The same pipeline is used for both Manual and Auto modes, ensuring that the experimental conditions differ in how the authoring specification is produced rather than in how the final animation is generated.

The synthesis process combines object-level motion with local stroke deformation. At the object level, affine transformations such as translation, rotation, and scaling follow the authored trajectory. At the stroke level, a separate deformation process introduces non-rigid motion such as bending and flexing. The resulting displacement therefore combines the coarse motion specified by the author with locally generated deformation, allowing the animation to remain expressive while preserving the intended object-level movement.

Because directly optimizing vector control points can introduce temporal instability, the synthesis pipeline additionally applies temporal regularization and an anchoring objective that encourages the generated geometry to remain consistent with the original sketch. The motion representation uses a hybrid Bernstein–Fourier formulation to balance flexible motion generation with temporal smoothness. These components stabilize the generated sequence while preserving the authored trajectory and the visual identity of the input sketch.

From the perspective of the mixed-initiative interface, the key property of this synthesis stage is that it is downstream of the authoring decisions. Once the user finalizes the object decomposition and trajectories, both Manual and Auto conditions invoke the same synthesis process. The technical details of the optimization procedure are provided in Appendix A.

\section{Preliminary Evaluation Study}
We conducted an exploratory within-subject study to examine how step-wise AI assistance changes the process of authoring multi-object sketch animations. Beyond measuring speed and output quality, we were interested in three linked questions: whether delegating intermediate decisions reduces authoring effort, how users remain involved when inspecting and correcting AI-generated proposals, and whether the resulting animations continue to align with the specified motion.

We therefore investigated the following research questions:

\begin{itemize}
    \item \textbf{RQ1 (Authoring Efficiency).} How does step-wise AI assistance affect the time and perceived effort required to author a multi-object sketch animation compared with direct manual authoring?
    
    \item \textbf{RQ2 (Delegation and Control).} How do users allocate work between accepting, directly repairing, and regenerating AI-proposed intermediate results, and how do these interactions shape their perceived control over the authoring process?
    
    \item \textbf{RQ3 (Perceived Outcome Effectiveness).} How do Manual and AI-assisted authoring differ in users' perceptions of motion-intent alignment and satisfaction with the resulting animation?
\end{itemize}

Together, these questions examine AI assistance not simply as a replacement for manual authoring, but as a process of delegating consequential creative decisions while retaining opportunities for human verification and intervention.

\subsection{Study Design and Conditions}

We used a within-subject design with two authoring conditions: \textbf{Manual} and \textbf{Auto}. Both conditions used the same sketches, motion descriptions, intermediate representations, and animation synthesis pipeline. They differed only in how object decomposition and motion trajectories were initially constructed.

In the \textit{Manual} condition, participants constructed the intermediate representation themselves. After selecting the assigned sketch and motion description, they created semantic object groups, assigned sketch strokes to the corresponding objects, and specified a trajectory for each animated object. Once satisfied with the specification, participants rendered the animation.

In the \textit{Auto} condition, participants began with the same sketch and motion description. OmniScribble generated the object decomposition and trajectory specifications sequentially. After each generated proposal, the system paused at a verification checkpoint. Participants could Accept the proposal, Edit Manually to modify the current representation directly, or Ask AI to Revise and optionally provide natural-language feedback. Once the intermediate representation was finalized, participants rendered the animation.

Thus, the two conditions differed in who initially constructed the intermediate decisions, while keeping the representation and downstream synthesis process constant. This design isolates the interactional effect of AI delegation from differences in the underlying animation technology.

\paragraph{Task materials and counterbalancing.}

We prepared five $256\times256$ multi-object vector sketches, one of which was reserved for tutorial and practice; the remaining four were used in the recorded study. One sketch was used only for tutorial and practice; the remaining four were used in the recorded study. Each participant completed four recorded authoring tasks: two in Manual mode and two in Auto mode, using different experimental sketches in each condition.

Sketch assignments were balanced across conditions so that each experimental sketch appeared equally often in Manual and Auto mode. The order of the two authoring conditions was also counterbalanced: half of the participants completed Manual before Auto, while the other half completed Auto before Manual. This design reduced potential confounds between authoring mode, task order, and sketch-specific difficulty.

\subsection{Participants and Apparatus}

\paragraph{Participants.}

We recruited 8 participants (4 female, 4 male; ages 20--25) through personal and university networks. Most participants were university students who had prior experience with generative AI tools and sketch-based applications. Their familiarity with these tools provided basic experience with AI-assisted creation and sketch-based interaction, which were relevant to the tasks in our study. None of the participants had prior experience with OmniScribble.

Before the study, participants received a brief introduction to the OmniScribble interface and completed a practice task to familiarize themselves with the main interactions. The practice task covered sketch selection, object construction, trajectory specification, and revision. The practice examples were different from those used in the main study. Participants were then asked to create animations that matched the specified motion using both Manual and Auto authoring modes.

\paragraph{Apparatus.}

The study was conducted in a quiet room using a desktop workstation and our browser-based sketch-animation prototype. The prototype was consistently configured for all participants, ensuring the same visual language model (VLM) setup, GPU-accelerated rendering pipeline, and interface components. Participants engaged with the same interface in both Manual and Auto modes. The two modes differed only in how object decomposition and motion trajectories were specified: in Manual mode, participants directly specified these components, while in Auto mode, the system generated initial proposals for participants to review and revise.

\subsection{Procedure}

Prior to the study, participants provided informed consent, completed a demographics questionnaire, and practiced both Manual and Auto modes using a tutorial sketch. Each participant then completed four recorded tasks, consisting of two tasks per mode. Mode order was counterbalanced across participants, and sketches were balanced so that each assigned drawing appeared equally in both conditions across the cohort.

For each task, participants reviewed the sketch and motion prompt, then initiated the session by pressing ``Start Task'' to begin the authoring timer. In the Manual condition, participants constructed semantic objects, assigned strokes, and defined motion trajectories by hand. In the Auto condition, the system generated initial proposals for object decomposition and trajectories, pausing after each stage for user verification. Participants evaluated each proposal by choosing to accept it, modify it directly, or request a regeneration.

Once the intermediate specification was finalized, participants clicked ``Render'' to stop the authoring timer. Rendering latency, questionnaire completion, and animation playback were therefore excluded from Authoring Completion Time. Participants then viewed the generated animation before proceeding to the next assigned task.

After completing both tasks within a condition, participants completed the condition-level questionnaire, including Task Ease, Perceived Control, Intent Match, Satisfaction, and Raw NASA-TLX. They were instructed to reflect on their overall experience across the two tasks completed in that authoring mode. Participants then transitioned to the alternate condition and completed two new sketches followed by the same questionnaire.

After all four recorded tasks were completed, participants took part in a semi-structured interview comparing their experiences across Manual and Auto modes. The interview focused particularly on their reasons for accepting, directly repairing, or regenerating AI proposals, as well as their perceptions of control, waiting time, and division of work between themselves and the AI.

\subsection{Measures and Analysis}

We structured the evaluation around three research questions concerning authoring effort (RQ1), delegation and control (RQ2), and perceived outcome alignment (RQ3).

\paragraph{RQ1: Authoring effort.}
We measured Authoring Completion Time as the elapsed time between ``Start Task'' and ``Render.'' This metric captured the interaction time required to produce a finalized animation specification, including time spent inspecting proposals, editing strokes or trajectories, and requesting
regenerations in the Auto condition. Animation synthesis latency, questionnaire completion, and playback duration were excluded.

Because each participant completed two tasks in each condition, task-level completion times were averaged within participant and condition before statistical comparison. This yielded one participant-level completion-time observation for Manual and one for Auto, avoiding the treatment of repeated tasks from the same participant as independent observations.

Subjective workload was assessed once after both tasks in each condition using the unweighted Raw NASA-TLX (RTLX) across its six subscales. Because the Performance item was collected with higher values indicating greater perceived success, we reverse-scored this dimension before computing RTLX so that higher values consistently represented greater workload. Task Ease was assessed on a 7-point Single Ease Question after both tasks in each condition.

\paragraph{RQ2: Delegation and control.}
To examine how participants interacted with AI-generated intermediate representations, we logged actions at the object and trajectory verification checkpoints. For each proposal, the logs recorded immediate acceptance, direct manual repair, and AI regeneration requests, including sequences that combined multiple correction strategies. These behaviors were summarized using checkpoint-level frequencies and participant-level intervention patterns.

Perceived Control was assessed once after both tasks in each condition on a 7-point scale. Post-study semi-structured interviews provided qualitative context for why participants accepted, repaired, or regenerated particular proposals and how they perceived the division of labor between themselves and the AI.

\paragraph{RQ3: Perceived outcome alignment.}
After completing both tasks in each condition, participants provided condition-level 7-point ratings of Intent Match and Satisfaction. Intent Match captured how well the resulting animations reflected the specified motions, whereas Satisfaction captured participants' overall evaluation of the generated outcomes. We analyzed these measures separately rather than
combining them into a composite score.

Given the exploratory nature of the study and the small sample size ($N=8$), we emphasize participant-level patterns, medians, interquartile ranges, and effect magnitudes. Exact Wilcoxon signed-rank tests were used for
the two primary quantitative endpoints, Authoring Completion Time and RTLX. Task Ease, Perceived Control, Intent Match, and Satisfaction are reported descriptively. Intervention strategies are likewise summarized descriptively because of the small number of observed correction events.

\section{Results}
We first examine whether AI assistance changed authoring effort, then analyze how participants interacted with AI-generated intermediate representations, and finally examine perceived outcomes. Because this was a small exploratory study, we focus on the magnitude and consistency of within-participant patterns rather than treating statistical significance as the primary evidence.

\subsection{RQ1: AI Assistance Reduced Authoring Effort}

AI-assisted authoring substantially reduced the effort required to construct a multi-object sketch animation. This pattern was evident in both objective task completion time and participants' subjective assessments of workload and task ease.

\paragraph{Authoring time.}
AI assistance substantially reduced authoring time, from a median of 232.68s (IQR = 211.09--261.06) in Manual mode to 71.68s (IQR = 61.67--84.40) in Auto mode (Fig.~\ref{fig:completion_time}).
All participants were faster in Auto mode, with a median within-participant reduction of 71.44\% (IQR = 56.75--75.75\%) and a median absolute reduction of 169.92s. An exact Wilcoxon signed-rank test indicated a
difference between conditions ($W=0$, $p=.0078$,
$\vert r_{rb}\vert=1.00$).

The reduction was consistent across both authoring stages. Object construction
decreased from a median of 85.52s in Manual mode to 24.93s in Auto mode, corresponding to a median within-participant reduction of 70.27\%.
Trajectory specification decreased from 157.87s to 46.20s, corresponding to a 71.39\% median within-participant reduction. Auto-mode trajectory times include the time participants spent inspecting and, when necessary, editing
or regenerating AI proposals, thereby capturing the full verification and correction workflow.

% \paragraph{Stage-level efficiency.}
% Time savings occurred across both intermediate authoring stages. For object construction, participant-level median time dropped from 85.52\,s (IQR = 73.74--92.76) in Manual mode to 24.93\,s (IQR = 23.42--26.50) in Auto mode, representing a 70.27\% reduction. Trajectory specification showed a parallel decline from 157.87\,s (IQR = 133.28--169.36) to 46.20\,s (IQR = 35.72--60.66), representing a 71.39\% reduction. Notably, the Auto trajectory times include additional interaction when participants chose to edit or regenerate an AI proposal. Thus, even when verification and correction were included in the authoring time, the AI-assisted workflow remained substantially faster than constructing the corresponding trajectories manually.

\begin{figure}[t!]
    \centering
    \includegraphics[width=0.8\columnwidth]{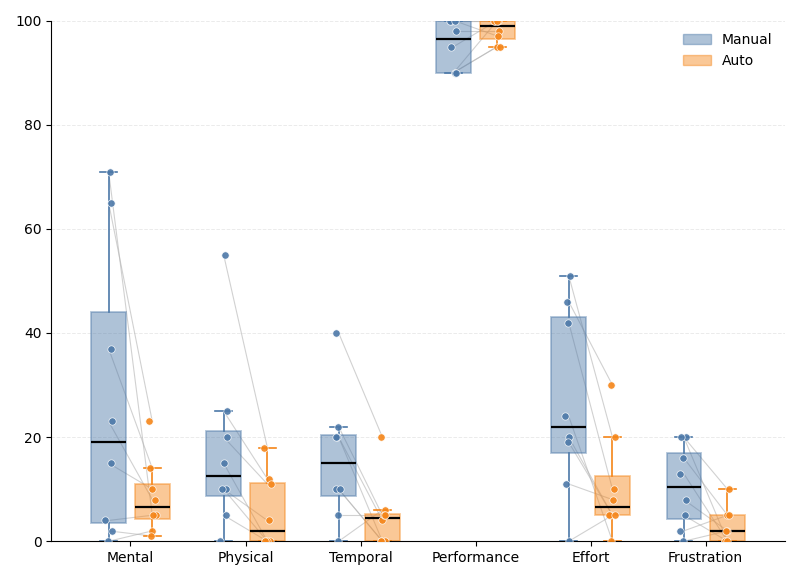}
    % \vspace{-5mm}
    \caption{NASA-TLX workload ratings under Manual and Auto authoring conditions ($N=8$). Box-and-whisker plots summarize the six workload dimensions, with individual participant responses overlaid. Higher values indicate greater perceived workload; Higher values indicate greater perceived workload; Performance ranges from Failure (0) to Perfect (100).}
    \Description{Box-and-whisker plots comparing NASA-TLX workload ratings between Manual and Auto authoring for eight participants. The six workload dimensions are Mental Demand, Physical Demand, Temporal Demand, Performance, Effort, and Frustration. Individual participant responses are overlaid on the distributions. Auto authoring generally shows lower workload ratings than Manual authoring for Mental Demand, Physical Demand, Temporal Demand, Effort, and Frustration, while Performance ratings are near the upper end of the scale in both conditions.}
    \label{fig:nasa-tlx}
\end{figure}

\paragraph{Perceived workload.}
The reduction in completion time was accompanied by lower subjective workload, with the median Raw NASA-TLX decreasing from 15.08 (IQR = 9.50--21.92) in Manual mode to 5.17 (IQR = 2.38--7.17) in Auto mode. Seven out of eight participants reported a reduced workload with AI assistance, confirmed by a paired Wilcoxon signed-rank test ($W=1$, $p\approx.016$). Notable reductions were seen in Mental Demand and Effort (Fig.~\ref{fig:nasa-tlx}).

\paragraph{Task ease.}
Participants found AI-assisted authoring easier, with Task Ease increasing from a median of 5.5 (IQR = 5.0--6.25) in Manual mode to 7.0 (IQR = 6.0--7.0) in Auto mode. Five rated Auto mode as easier, and no participant found it more difficult.

Taken together, these findings indicate substantial efficiency gains with AI-assisted authoring, with reductions in both completion time and perceived workload while retaining human verification within the workflow.

\subsection{RQ2: Delegation and Control}

The interaction logs show that participants did not uniformly accept or reject AI assistance. Instead, they used the checkpoint actions selectively, with most proposals accepted immediately and interventions concentrated in trajectory generation.

Across 16 AI-assisted tasks, participants encountered 32 verification checkpoints, evenly split between object construction and trajectory generation. They accepted 27 (84.4\%) immediately; five (15.6\%) involved intervention across five of eight participants (Tab.~\ref{tab:rq2-pathways}).

\paragraph{Object proposals required minimal intervention.}
Participants accepted 15 of 16 object proposals immediately (93.8\%). Only one proposal required manual stroke editing, and none prompted regeneration. The system-generated object decomposition proved sufficient to proceed directly to trajectory planning in nearly all cases.

\paragraph{Interventions concentrated on trajectory planning.}
Trajectory proposals prompted four interventions out of 16 cases (25.0\%), making this stage the primary locus of human--AI negotiation. Correction pathways varied: one trajectory was directly edited, one was regenerated twice, and two combined initial regeneration with subsequent manual editing. All regeneration requests across the entire study occurred at this stage.

\paragraph{Intervention increased authoring time.}
Immediately accepted trajectories required a median of 36.7~s (IQR = 33.2--38.2), while those involving repair or regeneration rose to a median of 92.7~s (IQR = 79.0--107.0). Given the small number of intervention cases, we treat this comparison descriptively. Nevertheless, intervention cases were associated with substantially longer elapsed trajectory-authoring times.

\paragraph{Perceived control remained high.}
Perceived Control remained high in both conditions. Median Perceived Control
was 5.5 (IQR = 5.0--7.0) in Manual mode and 6.0
(IQR = 5.75--7.0) in the AI-assisted condition. Four participants reported higher control with AI assistance, three reported identical ratings, and one reported lower control.

These traces indicate selective delegation. Direct repair provided targeted geometric adjustment, whereas regeneration delegated broader choreographic revisions back to the model, confirming that the two correction mechanisms served complementary roles in the AI-assisted workflow rather than functioning as interchangeable alternatives.

\paragraph{Why participants accepted, repaired, or regenerated AI proposals}
Semi-structured interviews revealed that participant actions were governed by the interplay among perceived alignment, interaction costs, and uncertainty regarding automated regeneration.

\textbf{Accepting aligned or satisficing proposals.}
Participants accepted proposals that matched their mental visualization, while tolerating minor deviations when further correction offered limited benefit. They also embraced unexpected but plausible motions as creative alternatives, making acceptance a form of both confirmation and exploration.

\textbf{Direct repair for localized adjustments.}
Participants used direct repair for inexpensive, localized errors, such as missing or misassigned strokes and small trajectory deviations. Manual editing also avoided the risk of stochastic regeneration overwriting valid components.

\textbf{Regenerating global semantic divergences.}
Participants preferred regeneration for broader trajectory mismatches that were difficult to reconstruct manually but easy to express through text, such as reversed direction or misaligned velocity. Targeted feedback therefore offered a more efficient way to specify global choreographic changes.

\begin{table}[t]
\centering
\caption{
Observed verification pathways across the 32 AI-generated intermediate
representations. Each row represents a mutually exclusive sequence leading
to a finalized intermediate representation.
}
\Description{Observed verification pathways for 32 AI-generated intermediate representations, divided equally between object proposals and trajectory proposals. The table lists four possible sequences: proposal followed by acceptance; proposal followed by manual repair and acceptance; proposal followed by two regenerations and acceptance; and proposal followed by regeneration, manual repair, and acceptance. Of the 32 representations, 27 were accepted immediately, 2 involved manual repair, 1 involved two regenerations, and 2 involved regeneration followed by manual repair. Object proposals accounted for 16 cases and trajectory proposals for 16 cases.}
\label{tab:rq2-pathways}
\small
\begin{tabular}{lccc}
\toprule
\textbf{Verification pathway}
& \textbf{Object}
& \textbf{Trajectory}
& \textbf{Total} \\
\midrule

Proposal $\rightarrow$ Accept
& 15
& 12
& 27 \\

Proposal $\rightarrow$ Manual Repair $\rightarrow$ Accept
& 1
& 1
& 2 \\

Proposal $\rightarrow$ Regenerate $\rightarrow$
Regenerate $\rightarrow$ Accept
& 0
& 1
& 1 \\

Proposal $\rightarrow$ Regenerate $\rightarrow$
Manual Repair $\rightarrow$ Accept
& 0
& 2
& 2 \\

\midrule
Total
& 16
& 16
& 32 \\

\bottomrule
\end{tabular}
\end{table}

\subsection{RQ3: Perceived Outcome Alignment Remained High Across Authoring Modes}

Despite the substantial reduction in authoring effort, participants rated the resulting animations positively in both conditions. Ratings remained near the upper end of the scale, although the small sample and concentration of scores at high values limit strong condition-level conclusions.

\begin{figure}[t]
    \centering
    
    \begin{subfigure}[t]{0.49\textwidth}
        \centering
        \includegraphics[width=\linewidth]{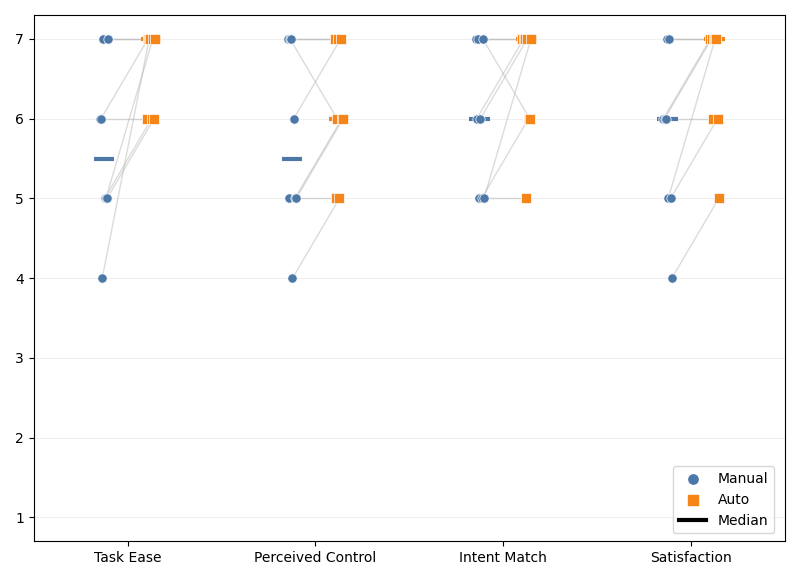}
        \caption{Participant-level subjective ratings.}
        \label{fig:subjective_ratings}
    \end{subfigure}
    \hfill
    \begin{subfigure}[t]{0.49\textwidth}
        \centering
        \includegraphics[width=\linewidth]{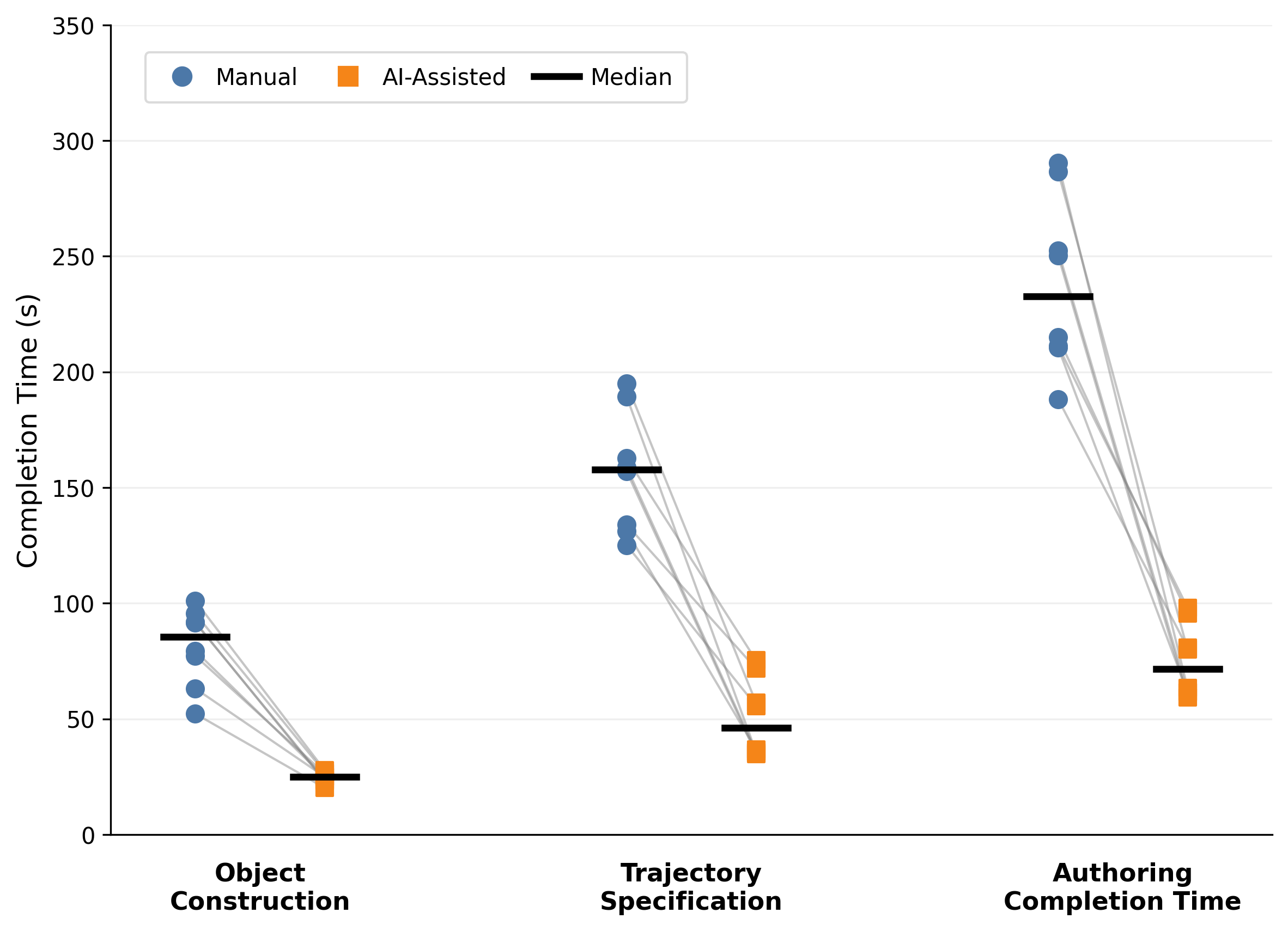}
        \caption{Participant-level authoring completion time.}
        \label{fig:completion_time}
    \end{subfigure}
    
    \caption{
    Participant-level comparison between Manual and Auto authoring.
    (a) Condition-level subjective ratings collected after participants completed both tasks in each authoring mode. Each connected pair represents one participant ($N=8$), and horizontal bars indicate condition medians. Task Ease ranges from 1 (Very Difficult) to 7 (Very Easy); the remaining measures range from 1 (Strongly Disagree) to 7 (Strongly Agree). Higher values indicate more favorable ratings.
    (b) Authoring completion time. Circles and squares indicate Manual and Auto observations, respectively. Horizontal bars indicate condition medians. All participants had lower mean completion times in Auto mode.
    }
    \Description{Participant-level comparison of Manual and Auto authoring for eight participants. Panel (a) shows paired ratings for Task Ease, Perceived Control, Intent Match, and Satisfaction; each participant's Manual and Auto ratings are connected, with horizontal lines indicating condition medians. Panel (b) shows authoring completion times for Object Construction, Trajectory Specification, and the overall Authoring Completion Time; paired observations are connected and condition medians are marked. Auto authoring has lower completion times for all participants and generally favorable subjective ratings.}
    \label{fig:user_study_results}
\end{figure}

\paragraph{Motion-intent alignment.}
Intent Match had a median of 6 (IQR = 5--7) in Manual mode and 7 (IQR = 6--7) with AI assistance. Four participants rated the AI-assisted result higher, three provided identical ratings across conditions, and one rated Manual mode higher
(Fig.~\ref{fig:subjective_ratings}). Given the small sample and concentration of ratings toward the upper end of the scale, we interpret
these results descriptively rather than as evidence of equivalence.

\paragraph{Satisfaction.}
Satisfaction descriptively favored AI assistance. Median Satisfaction increased from 6 (IQR = 5.0--6.25) in Manual mode to 7 (IQR = 6.0--7.0) in Auto mode. Five participants reported higher Satisfaction with AI assistance, while three provided identical ratings and none favored Manual mode.

% Together, the reductions in authoring time and workload with AI assistance were not accompanied by an apparent decline in participants' evaluations of the resulting animations. Given the exploratory sample and the concentration of subjective ratings near the upper end of the scale, these findings characterize the observed outcomes in our controlled tasks rather than establishing equivalence between the two authoring modes.

\section{Discussions}
% Our study suggests that AI assistance in sketch-animation fundamentally redistributes authoring labor between the user and the system. Participants completed the AI-assisted tasks substantially faster and with lower perceived workload, while perceived control remained high and the resulting animations were rated as strongly aligned with the specified motion in both conditions. Taken together, these findings point to a design space between fully manual authoring and opaque end-to-end generation: consequential creative decisions can be delegated to AI while keeping them visible and open to user intervention.

Our study suggests that AI assistance in sketch animation redistributes authoring labor between users and the system. Participants completed AI-assisted tasks substantially faster and with lower perceived workload, while perceived control remained high and resulting animations were rated as strongly aligned with the specified motion. Rather than replacing authorial decisions through opaque end-to-end generation, OmniScribble delegates consequential decisions while keeping intermediate representations visible and open to intervention.

Our findings indicate that AI assistance in OmniScribble not only reduces the time spent on authoring but also changes how users allocate their effort during the creative process. While the AI significantly speeds up the construction of object representations and trajectories, participants still have the opportunity to inspect and revise their work. Semi-structured interviews revealed that the use of AI varies among users; they often evaluate whether an AI proposal is suitable, determine if fixing a mismatch is more efficient for them personally, or decide whether to return the issue to the AI. We explore these behaviors in relation to mixed-initiative creative authoring and its broader implications.

% Our findings suggest that the value of AI assistance in OmniScribble lies not simply in reducing authoring time, but in changing how users allocate effort throughout the creative process. AI-assisted authoring substantially reduced the time required to construct object representations and trajectories, while participants still retained opportunities to inspect and revise intermediate results. At the same time, the semi-structured interviews show that delegation was neither uniform nor unconditional. Participants continually decided whether an AI proposal was sufficient, whether a mismatch was cheaper to repair themselves, or whether the problem should be delegated back to the AI. We discuss these behaviors as implications for mixed-initiative creative authoring, sustained use, and broader application contexts.

\subsection{Correction as Dynamic Allocation of Human--AI Effort}

The correction behavior observed in our study suggests that participants did not treat AI proposals as simply ``correct'' or ``incorrect.'' Instead, their decisions to accept, directly repair, or regenerate a proposal reflected a trade-off between the quality of the current result and the perceived cost, uncertainty, and expected benefit of further correction. This helps explain why most proposals were accepted immediately, while five of eight participants nevertheless intervened at least once during their AI-assisted tasks. High acceptance therefore did not necessarily indicate passive use; rather, participants appeared to intervene selectively when the expected value of correction justified the additional effort.

\paragraph{Acceptance based on sufficiency and serendipity.}
Participants accepted proposals that matched their mental image of the caption, but exact alignment was not always required. Authors frequently accepted imperfect proposals when the result was practical enough and further refinement was not worth additional waiting time. In several instances, participants also expected the downstream synthesis stage to smooth out minor trajectory flaws. In other cases, unexpected paths were adopted because they offered appealing alternatives, turning intermediate proposals into sources of creative variation.

\paragraph{Direct repair to protect progress.}
Direct manipulation was preferred for localized spatial errors. Reassigning a few strokes or moving isolated path waypoints took less time than waiting for new system proposals. Direct repair also served as a risk-management strategy. Because full regeneration could discard well-formed components, manual editing allowed users to fix small errors while protecting satisfactory parts of the layout.

\paragraph{Regeneration for high-level semantic adjustments.}
Regeneration was favored when errors involved overall motion characteristics that were easier to describe with words than to draw manually, such as reversing trajectory directions or modifying speeds. However, because each regeneration introduced the risk of altering functional parts of the scene, participants often requested an initial regeneration to fix global motion and then switched to manual repair for final adjustments.

\subsection{Designing for Sustained Mixed-Initiative Authoring}

Our interviews further suggest that the long-term usefulness of AI assistance depends not only on generation quality, but also on how quickly the system responds, how easily users can preserve previous work, and whether direct manipulation remains available when it is the cheaper or more satisfying form of interaction.

\paragraph{Latency shifts the division of creative labor.}
Participants tolerated delays ranging from 20 seconds to several minutes, indicating that acceptable latency depends on authoring context. Waiting disrupted momentum when participants had a concrete hypothesis to test, but brief pauses could also provide opportunities to reconsider the composition. Latency therefore shapes task allocation: slower generation encourages users to handle simple corrections manually, whereas faster generation supports open-ended experimentation. Participants nevertheless preferred direct manipulation for localized, deterministic adjustments.

\paragraph{Non-destructive regeneration protects progress.}
Regeneration can unnecessarily discard viable work when a proposal contains both useful elements and minor errors. Future systems should support partial regeneration, allowing users to lock approved objects or path segments while regenerating only unresolved regions. Version history and multi-candidate previews could further support the incremental correction patterns observed in our study.

\paragraph{Direct manipulation preserves creative agency.}
Participants associated creative ownership with guiding consequential choices, such as adjusting trajectories and confirming scene decomposition, rather than performing every mechanical action manually. Direct manipulation therefore provides a means to retain authorship within an automated workflow. Rather than separating manual and automated authoring into distinct modes, co-creative systems should support switching between them at the level of individual actions, enabling delegation of structural work while preserving tactile control over expressive details.

\subsection{Application Opportunities and Generalizability}

\begin{figure}[!t]
    \centering
    \includegraphics[width=0.75\textwidth]{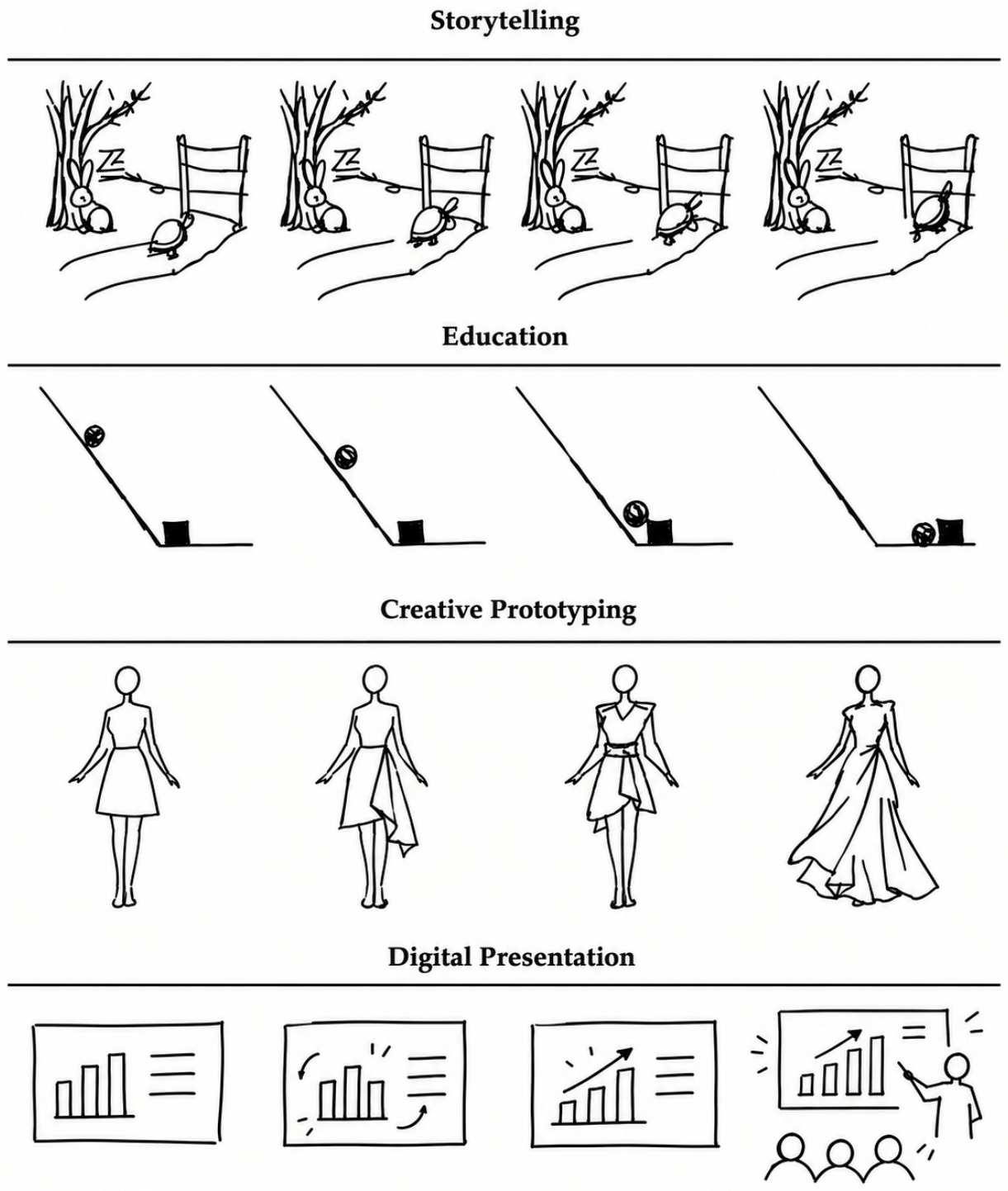}
    \vspace{-1cm}
    \caption{Illustrative applications of OmniScribble across storytelling, education, creative prototyping, and digital presentation. The examples show how static multi-object sketches can be transformed into animated scenes for communication, design, and learning.}
    \Description{A figure showing various applications of OmniScribble, including storytelling, education, creative prototyping, and digital presentation, with examples of static sketches transformed into animated scenes.}
    \Description{Illustrative application scenarios for OmniScribble, organized into four rows: Storytelling, Education, Creative Prototyping, and Digital Presentation. Each row shows a sequence of static sketch frames representing progressive motion or transformation. The examples illustrate how multi-object sketches could be animated for narrative scenes, instructional or physics demonstrations, fashion and design exploration, and presentation graphics.}
    \label{fig:applications}
\end{figure}

Participants' responses also point to application contexts where rapid generation and editable intermediate representations could be useful, as illustrated in Fig.~\ref{fig:applications}. We frame these as application opportunities rather than demonstrated outcomes, since our study evaluated short authoring tasks in a controlled setting.

\paragraph{Storytelling and children's narratives.}
For open-ended storytelling, unexpected AI-generated motions can support creative exploration. Users can delegate initial motion synthesis and selectively refine paths by hand, balancing narrative intent with serendipitous system suggestions.

\paragraph{Educational and explanatory animation.}
Mixed-initiative authoring could help educators create precise instructional animations without manually constructing complex scenes. OmniScribble provides rapid motion proposals while keeping intermediate representations editable, allowing instructors to verify and refine details such as trajectories, timing, and causal relationships in physics examples.

\paragraph{Generalizing beyond sketch animation.}
These interaction patterns could extend to generative workflows with structured intermediate representations, such as UI wireframes, 3D scene graphs, and temporal video timelines. Across these domains, mixed-initiative systems can expose consequential decisions through checkpoints that support inspection, partial preservation, direct editing, and re-prompting rather than relying solely on end-to-end automation.

\subsection{Limitations}
\paragraph{System limitations.}

OmniScribble remains sensitive to the visual complexity and abstraction of input drawings. Highly sparse sketches lack sufficient visual evidence for the model to infer semantic boundaries reliably, while densely entangled strokes make manual stroke selection tedious and error-prone. Additionally, coarse object-level waypoints constrain expressive fidelity; complex behaviors remain difficult to control through path coordinates alone. Furthermore, user interaction strategies were tied to model performance and response latency. While subsequent synthesis regularizes local motion, future iterations require richer physical timing parameters, localized component locking, and partial regeneration to prevent destructive overwriting.

\paragraph{Study limitations.}

Our preliminary evaluation involved a small cohort ($N=8$). Although the paired within-subject design exposed consistent participant-level differences, the sample size limits statistical generalization. Moreover, the protocol evaluated predefined tasks and motion prompts. This controlled setting supported systematic comparison across modes, but constrained open-ended creative exploration, meaning that intent match ratings reflected compliance with assigned instructions rather than unconstrained personal expression. Finally, the study examined short-term interactions without an opaque, end-to-end generative baseline. As a result, our findings contrast manual construction directly against inspectable delegation, but do not isolate the specific causal impact of intermediate checkpoints against single-shot automation. Evaluating long-term trust calibration, deploying the system in open-ended creative environments, and incorporating blind external assessments of animation quality represent essential next steps for validating inspectable mixed-initiative authoring.

\section{Conclusion}

We presented OmniScribble, a mixed-initiative system that structures multi-object sketch animation through inspectable and repairable intermediate representations. By exposing object decomposition and motion trajectories as editable artifacts, the interface allows authors to either construct scene elements directly or evaluate AI-generated proposals through acceptance, direct repair, and language-guided regeneration. This checkpoint-based mechanism enables users to flexibly adjust the division of labor across authoring stages. Our within-subject study demonstrated that AI assistance yielded a median within-participant reduction in authoring time of 71.4\% and significantly lowered subjective workload, while perceived control and motion-intent alignment remained high across conditions. Semi-structured interviews confirmed that participants intervened selectively, adopting direct repair to resolve localized geometric errors and requesting regeneration for broader choreography adjustments. These results indicate that successful human-AI co-creation depends on making automated decisions visible, editable, and reversible, enabling creators to streamline execution while retaining control over consequential creative outcomes.

\bibliographystyle{ACM-Reference-Format}
% \balance
\bibliography{ref}

% \newpage

% \appendix
% \input{appendix/methodology2}
% \input{appendix/results}

\end{document}